# Tunable giant exchange bias in single-phase rare earth-transition metal intermetallics $YMn_{12-x}Fe_x$ with highly homogenous inter-sublattice exchange coupling

Yuanhua Xia[1]*, Rui Wu[1]*, Shunquan Liu[1], Honglin Du[1], Jingzhi Han[1], Changsheng Wang[1], Xiping Chen[3], Lei Xie[3], Y.C. Yang[1], Jinbo Yang[1,2†]

[1]State Key Laboratory for Mesoscopic Physics, School of Physics, Peking University, Beijing 100871

[2]Collaborative Innovation Center of Quantum Matter, Beijing, 100871, P.R. China

[3]Institutes of Nuclear Physics and Chemistry, China Academy of Engineering Physics, Mianyang, 621999, China

**In this paper, we have found a family of intermetallic compounds $YMn_{12-x}Fe_x$ (x = 6.6-8.8) showing a bulk form of tunable giant exchange bias effect which arises from global interactions among ferromagnetic (FM) and antiferromagnetic (AFM) sublattices but not the interfacial exchange coupling or inhomogeneous magnetic clusters. A giant exchange bias with a loop shift up to 6.1 kOe has been observed in $YMn_{4.4}Fe_{7.6}$ compound with the strongest competing magnetic interactions. In a narrow temperature range, the exchange bias field shows a sudden switching off whereas the coercivity shows a sudden switching on with increasing temperature. This unique feature indicates that the inter-sublattice exchange coupling is highly homogenous, which can be perfectly interperated by our theoretical calculations.**

The exchange bias (EB), which usually refers to a shift of the hysteresis loop along the field axis in the exchange coupled ferromagnetic (FM)/antiferromagnetic (AFM) systems, is a phenomenon firstly discovered in Co/CoO nanoparticles in 1956 [1]. The EB effect has significant impacts on the technological applications of data storage products, spintronic devices, permanent magnets, and many other devices [2-8].

Extensive research has led to the notion that EB must originate from uncompensated interfacial spins that are pinned in the AFM and cannot be reversed by external field after a field cooling (FC) procedure through the Néel temperature ($T_N$) of the AFM [9,10]. After the discovery of this effect in Co/CoO nanoparticles, investigations of the EB effects have been mainly focused on a large number of heterogeneous structures such as magnetic bilayers, core-shell nanoparticles, and FM nanoparticles embedded in an AFM matrix compounds [6-13]. The EB effects have also been observed in single phase bulk oxides and alloys with competing exchange interactions which always result in magnetic phase separations or spin glass state in these systems [14-19]. For example, a zero field cooling (ZFC) EB (ZEB) effect has been realized in Ni-Mn-In and Mn-Pt-Ga Heusler alloys [15-17]. The super spin glass (SSG) phase and the FM inclusions embedded in ferrimagnetic (FIM) ordering matrix were proposed to play key roles in Ni-Mn-In and Mn-Pt-Ga, respectively. It is obvious that they are all structurally single-phased bulk materials, but with multiple magnetic phases, which can give rise to EB effect by exchange couplings at the interfaces of different magnetic phases. Besides, EB has also been reported in $YbFe_2O_4$ system with exchange interaction taking place at low temperature between FM $Yb^{3+}$ and FIM $Fe^{2+}/Fe^{3+}$ sublattices [20].

The rare-earth intermetallics have been proven to be a fertile research area due to its fascinating physical properties including hard magnetic properties, giant magnetostriction effect, giant magnetocarloric effect, etc. [21-23]. Among them, Mn is the only 3d magnetic transition metal that stabilizes binary compounds of $ThMn_{12}$-type structure (space group I4/mmm), leading to the highest metal to rare-earth ratio in the rare-earth intermetallic [24]. In $RMn_{12}$ compounds, the Mn ions are coupled antiferromagnetically showing Néel temperatures around 100 K. The substitution of Fe for Mn in $RMn_{12-x}Fe_x$ promotes ferromagnetic ordering [25]. The rare earths occupy the 2a sites while Fe/Mn atoms occupy three nonequivalent sites: 8i, 8j and 8f (See supplementary Fig. S1), with strong site preferences with the 8i sites favoring Mn and the 8f sites favoring Fe atoms [26, 27]. The magnetic arrangements of the 3d sublattices are antiferromagnetic for x < 6; then transform progressively to more complex configurations of "FM + AFM" for 6 < x < 9; and finally to a purely ferromagnetic one for the iron-rich compounds (x = 9). Therefore, we consider the compound of $RMn_{12-x}Fe_x$ as a single-phase magnet that consists of two kinds of magnetic sublattices: one is AFM structure and another is FM structure, mimicking that of an artificial FM/AFM superlattice. Thus, the EB effect could be realized in this system and could be easily tuned by the concentration of Fe element.

In this letter, we found the existence of EB effect in rare earth transition metal compounds $YMn_{12-x}Fe_x$ (x = 6.0-8.8) bulk alloys where the pinning phenomenon is caused by highly homogenous global interaction between FM-AFM sublattices. In $YMn_{4.4}Fe_{7.6}$, with the strongest competing magnetic interactions, the relatively high ordering temperature and low magnetic anisotropy in the AFM site lead to a very sharp switching characteristic of the EB effect far below the Neel temperature $T_N$. Based on the inter-sublattice interactions, a theoretical model was established and was found to corroborate well with the experimental results.

The polycrystalline $YMn_{12-x}Fe_x$ (x = 6.0-8.8) bulk samples were prepared by arc melting of 99.9% pure materials in a purified argon atmosphere. An excess rare earth and manganese were added to compensate for their losses during melting. Then the ingots were annealed in an evacuated and sealed silica tube at 1000 ℃ for 5 days. The crystal and magnetic structures of these compounds were determined using the X-ray (XRD) diffraction with Cu-Kα radiation and neutron diffraction at various temperatures. The refinement of XRD and neutron diffraction indicates that the samples (x = 6.0-8.6) are a single phase with tetragonal $ThMn_{12}$-type structure while $YMn_{3.2}Fe_{8.8}$ contains a small amount of $Y(Fe,Mn)_2$ phase. Magnetization was measured using magnetic property measurement system (MPMS-7) and the physical property measurement system (PPMS).

To characterize the magnetic properties of $YMn_{12-x}Fe_x$ compounds, the temperature dependence of the magnetization (*M*-*T*) was measured. Fig. 1 shows the magnetic phase diagram of $YMn_{12-x}Fe_x$ ($6.0 \leq x \leq 8.8$) under a magnetic field of 1 kOe according to the *M*-*T* data. It can be seen that most of the samples have three transition temperatures corresponding to Curie temperature ($T_C$, blue circle), Neel temperature ($T_{N,}$ red triangle), and spin freezing temperature ($T_f$, black square), respectively. The $T_C$ monotonously increases whereas $T_N$ decreases with the increase of Fe content. Along with the temperature climbing, the samples experienced paramagnetic, ferromagnetic, antiferromagnetic, and glassy magnetic state changes for x > 7.6, and paramagnetic, antiferromagnetic, ferromagnetic, and glassy magnetic state changes for x < 7.6. Therefore, the ferromagnetic and antiferromagnetic exchange coupling effect becomes stronger at low temperature. The strongest competence between ferromagnetic and antiferromagnetic interactions can be expected at the crossing point of the $T_C$ and $T_N$ curves, which corresponds to the highest freezing temperature $T_f$. This suggests that the competence between ferromagnetic and antiferromagnetic interactions may lead to a large EB effect for the x = 7.6 sample.

The Fe content dependence of EB effect in the $YMn_{12-x}Fe_x$ alloys (6.0 < x < 8.8) was investigated. Table I listed the EB fields ($H_E$) and coercivities ($H_C$) of the $YMn_{12-x}Fe_x$ alloys (6.0 < x < 8.8) under FC conditions, where $H_E$ and $H_C$ are defined as $H_E = -(H_L + H_R)/2$ and $H_C = -(H_L - H_R)/2$, respectively, with $H_L$ and $H_R$ being the left and right coercive fields. It was found that all of the samples show EB effect under FC condition. After 1 kOe FC procedure from the room temperature, the $H_E$ value increases with the decrease of Fe content and reaches a maximum of about 5.96 kOe at x = 7.6, then decreases with further decrease of Fe content. The maximum $H_E$ corresponds to the strongest competing exchange interaction between the Fe and Mn sublattices. In addition, in the samples with x = 7.0 and 6.6, spontaneous EB effect with large coercivities was found under the ZFC condition and little difference of coercivity $H_c$ was revealed between FC and ZFC conditions.

Fig. 2 (a) displays the typical ZFC and FC magnetization curves for $YMn_{4.4}Fe_{7.6}$ under applied field of 100 Oe. The ZFC curve exhibits two peaks at $T_N$ =163 K and $T_f$ = 130 K, and there is a bifurcation between the ZFC and FC curves around $T_f$. The antiferromagnetic ordering of Mn (8i) magnetic sublattice occurs at $T_N$ ~ 163 K. There is another magnetic transition around 143 K ($T_C$) due to the ferromagnetic ordering of magnetic moments at Fe (8f) and (8j) sites. Below 100 K, the ZFC magnetization drops with the decreasing temperature, indicating that the spontaneous interaction between Mn and Fe sublattices prefers an AFM configuration. As shown in Fig. 2(b), from the real part of ac susceptibility curve of $YMn_{4.4}Fe_{7.6}$, two distinct peaks can be observed (the arrows indicate the positions). One peak corresponds to the antiferromagnetic ordering temperature $T_N$ that does not change with frequency; the other peak corresponds to spin freezing temperature $T_f$ that moves with increasing frequency to high temperature region. This indicates that the material has the characteristics of a spin glass due to the competing interactions among different magnetic sublattices. What makes the system interesting is that the inter-sublattice coupling is relatively weak compared to the cooling field and can be manipulated by a FC process under moderate magnetic fields. As shown in Fig. 2 (a), after cooling the sample in 100 Oe from 300 K, the FC magnetization goes up instead of dropping down below ~125 K, indicating a parallel alignment of two magnetic sublattices after FC. Thus, the relative orientation between Mn and Fe magnetic sublattices can be effectively manipulated by the FC process. Fig. 2 (c) shows the *M-H* hysteresis loops at 5 K for $YMn_{4.4}Fe_{7.6}$ with various cooling fields. The FC *M-H* loops shift left along *H*-axis with positive fields and shift right along *H*-axis with negative magnetic fields, while the ZFC *M-H* loop exhibits nearly symmetric coercive fields. The EB field $H_E$ reaches a giant value of 6.1 kOe with a cooling field of 50 kOe, much larger than EB fields reported before for rare earth-based intermetallics.

We then studied the dependence of EB effect on the cooling field $H_{\mathrm{cool}}$ at 5 K. As shown in Fig.2 (d), $H_E$ increases rapidly with the increasing $H_{\mathrm{cool}}$ and nearly reaches its saturation value at $H_{\mathrm{cool}}$ = 1 kOe. With further increase of the cooling field, $H_E$ increases slightly and remains almost constant up to 50 kOe. This feature of cooling field dependence is different from that in the phase-separated oxides or spin glass systems where $H_E$ decays rapidly with high cooling magnetic fields due to the growth of the FM clusters or the melting of spin glass [14, 28-29]. This indicates that in $YMn_{12-x}Fe_x$, the metastable spin configuration is reasonably stable even against the applied magnetic field. The uniqueness of $YMn_{12-x}Fe_x$ suggests that the observed giant EB may not be an interfacial coupling type resulting from phase separation or spin glass, two major mechanisms for the EB effect in single phase compounds and alloys that have been reported so far.

To further investigate the giant EB effect, we studied the temperature dependence of EB under FC conditions. Fig. 3 shows the hysteresis loops of $YMn_{4.4}Fe_{7.6}$ alloy at various temperatures. The characteristics of the hysteresis loops reveal three types of changes with the increasing temperature: (I) in the range of 5-20 K, both branches of the hysteresis loop are coincides and show no significant variation with the temperature (in fig 3(a)); (II) in the range of 20-30K, the left branch of the loop shifts slightly to the left direction, while the right branch shifts quickly along x-axis to the right direction until a symmetric loop forms at about 30 K (in fig 3(a)); (III): both left and right branches gradually shift back to the original point (0,0) until they coincide with the temperature above 30 K (Fig. 3(b)). Under ZFC condition (see supplementary Fig. S2), when the temperature is less than 20 K, both branches of the hysteresis loop are coincides. With the increasing temperature, the coercivity first remains constant, then sharply rises to a maximum value at 25 K, and in the end decreases gradually.

To analyze the impact of temperature on the EB effect, the temperature dependence of $H_L$, $H_R$, $H_C$ and $H_E$ are plotted in Fig 4. In the temperature range of 5-20 K, both $H_L$ and $H_R$ take the value of about -6.1 kOe and remain almost constant. The values of $H_R$ alter abruptly from negative 5.8 kOe to positive 4.5 kOe in the range of 20-28 K, and then start to decrease with further increase of the temperature, while the $H_L$ decreases gradually after 25 K, and reaches zero at 150 K. In accordance to the changes of $H_L$ and $H_R$, $H_E$ first maintains at about 6 kOe at T < 20 K, then decreases rapidly to -113 Oe at about 32 K, and nearly disappears at the blocking temperature $T_B$ = 40 K. The spin freezing temperature $T_f$ (see fig. 2(a)) of about 120 K is well above the $T_B$, suggesting that the EB effect is not related to the spin glass phase in this system. The $H_C$ shows a tiny value close to zero at T < 20 K, and increases abruptly after 20 K and reaches the maximum of 5 kOe at 28 K, and then decreases slowly to close to zero at 150 K. The sharp changes of $H_C$ and $H_E$ at around 25 K are related to the high homogeneity of the pining in this system, in which the large EB effect originates from the exchange interaction between different magnetic sublattices.

This interpretation is supported by the negligibly small training effect in $YMn_{4.4}Fe_{7.6}$, where EB field decreases only 0.16 % during the first 5 consecutive field cyclings at 5 K after FC from 300 K with $H_{cool}$ = 50 kOe, just like what Fig 3(a) insert shows. In FM/AFM heterostructures, the training effect is originated from the heat activated fluctuations in the FM-AFM exchange coupling, which modify the magnetic interactions between differently coupled regions with each reversal [30-32]. For a system with high homogeneity, all AFM spins act like one macroscopic spin and show no fluctuation. Consequently, there can be only two states in this system, all AFM spins contribute to the EB or no AFM spins contribute to the EB. Thus, the training effect, a state that only a part of the AFM spins contribute to the EB, will be absent in this system. The negligibly small training effect well supports the proposed highly homogenous inter-sublattice exchange coupling mechanism for the EB in this system.

In order to establish a comprehensive picture of the above results, the $YMn_{12-x}Fe_x$ alloys were considered as a single-phase magnet that consists of two magnetic sublattices: one with AFM structure and another with FM structure. The spin configuration is therefore similar to that of an artificial FM/AFM superlattice. According to neutron diffraction data refinement, the magnetic structure of the $YMn_{4.4}Fe_{7.6}$ is presented in Fig 5(a). The result is similar to what has been described previously [24-25]. The magnetic interactions between Mn-Mn, Fe-Fe and Mn-Fe are AFM, FM and AFM, respectively. Fe and Mn atoms prefer to occupy 8f and 8i sites, respectively, while they randomly distribute on the 8j site. Due to the single phase character of $YMn_{4.4}Fe_{7.6}$, the FM/AFM interface can exist between different magnetic sublattices. It can be seen from Fig.5(a), $YMn_{4.4}Fe_{7.6}$ forms two sets of magnetic lattices: Mn atoms on 8i sites give rise to the AFM coupling magnetic sublattice (blue atoms); Fe atoms on 8f sites give rise to the FM coupling magnetic sublattice (red atoms). Mn and Fe atoms on 8j site form the "AFM/FM interface", which leads to AFM interaction between 8j-8i sublattices, and AFM or FM interaction between 8j and 8f sublattices with Mn or Fe occupying on 8j site, respectively. Due to the fact that 8j site can be occupied by Fe or Mn atoms evenly, the stoichiometry of Fe content directly affects Fe and Mn ratio at the interface, and subsequently influences the exchange coupling interaction between AFM and FM. With the increase of Fe content, the 8j sites (interface) becomes FM dominated and may lead to smaller $H_E$ and $H_C$ due to weaker pinning effect of the interface layers. This is confirmed by the results in Table I, where different degrees of EB effect were observed in the samples with different Fe contents.

According to the above discussion, each sublattice of $RMn_{12-x}Fe_x$ can be viewed as a single spin due to its high homogeneity. The generalized Meklejohn-Bean (M-B) model was adopted to explain the EB in $RMn_{12-x}Fe_x$ compounds, which consists of AFM (Mn atoms from 8i site and 8j site), FM (Fe atoms from 8f site and 8j site). The energy of the system is given by [33]

$$E = -J_{FM-AFM} M_{FM} M_{AFM} \cos(\beta - \alpha) - K_{FM} (M_{FM} \cos(\theta - \alpha))^2 - K_{AFM} (M_{AFM} \cos(\varphi - \beta))^2 - HM_{FM} \cos(\alpha) - HM_{AFM} \cos(\beta) \quad (1)$$

where $H$ and $M_{FM(AFM)}$ are externally applied magnetic field and saturation magnetization of FM(AFM) layers, respectively. $K_{FM(AFM)}$ is effective uniaxial magnetic anisotropy constant of FM(AFM) layer, $J_{FM-AFM}$ is the exchange coupling between the FM and AFM layers, while $\alpha$, $\beta$, $\theta$ and $\varphi$ are the azimuth angles of $M_{FM}$, $M_{AFM}$, uniaxial anisotropy axis of FM and uniaxial anisotropy axis of AFM with respect to the applied magnetic field direction, respectively, as the Fig. 5(b) shows.

As we know, the temperature dependence of EB is relevant to thermal instabilities of the AFM interfacial magnetization [34]. According to the Néel-Brown relaxation theory, the Néel relaxation time $\tau_N$ of the AFM magnetization should be

$$\tau_N = \tau_0 \exp\left( \frac{K_{AFM} V}{k_B T} \right) \quad (2)$$

Then, the contributions to the EB effect from increasing temperature $T$ in our experiment can be analogous to that from the decreasing antiferromagnetic anisotropy $K_{AFM}$ in our calculation model with the relation of $T \sim K_{AFM}^{-1}$.

According to Eqn. (1), the hysteresis loops with different $K_{AFM}$ are calculated and are plotted in Fig. 5(c). The obtained coercivities $H_C$ and the EB fields $H_E$ are shown as the function of $K_{AFM}^{-1}$ in Fig. 5 (d) and (e). In the calculation, we have assumed a collinear alignment of the anisotropy axes and the external field. All the parameters are dimensionless with $M_{FM}$ = 1 and $M_{AFM}$ = 0.5$M_{FM}$, $K_F$ = 0.1$M_{FM}$, $H_{max}$ = 2$M_F$, $J_{FM\text{-}AFM}$ = -4$M_{FM}$. The antiferromagnetic anisotropy $K_{AFM}$ changes between 500

$M_{FM}$ and 5 $M_{FM}$.

As can be seen from Fig. 5(d) and (e), the calculated results, especially the sudden switching on of the EB field and the sudden switching off of the coercivity at low temperatures, agree very well with the experimental data (Fig. 4). When the temperature is lower than 20 K, the anisotropy of the antiferromagnetic spins is so large that the AFM spins remain in their original configuration while the FM spins rotate with the external magnetic field. This will give a loop shift of the ferromagnetic spins due to the unidirectional pining from the antiferromagnetic spins. However, when the temperature is higher than 20 K, the antiferromagnetic spins begin to rotate with the ferromagnetic spins under the interfacial exchange coupling due to the decrease of antiferromagnetic anisotropy $K_{AFM}$ with the increasing temperature. Then, the bidirectional pining from the antiferromagnetic spins will give an enhancement in the coercivities of the ferromagnetic spins. The sharpness of the switching process in the experimental results indicates the homogeneity of the FM-AFM interface, corresponding to the special global inter-sublattice coupling in $YMn_{4.4}Fe_{7.6}$.

## Conclusions

Tunable giant EB effect in a family of rare earth-transition metal intermetallic compounds $YMn_{12-x}Fe_x$ (x = 6.6-8.8) was observed, resulting from a competing magnetic interaction among ferromagnetic (FM) and antiferromagnetic (AFM) sublattices. A maximum EB with a loop shift up to 6.1 kOe has been revealed in this single-phase bulk alloy $YMn_{4.4}Fe_{7.6}$. The EB field remains almost unchanged at temperatures below 20 K, but shows a sudden switching off in the temperature range of 25~30 K, where the coercivity shows a sudden switching on and a subsequent slow decrease. The calculating results of the theoretical model show an excellent agreement with the experimental result of this unique temperature dependence of EB. This indicates that the large exchange anisotropy originates from the highly homogenous exchange interaction between Fe-rich and Mn-rich sublattices.

**ACKNOWLEDGMENTS :**This work was supported the National Natural Science Foundation of China (Grant Nos. 51171001, 51371009, 50971003 and 11504348). The authors thank the Key Laboratory of Neutron Physics of CAEP for providing the neutron beam time.

*Y. Xia and R. Wu contributed equally to this work.

†Corresponding author: jbyang@pku.edu.cn

**REFERENCES**

1. W. H. Meiklejohn and C. P. Bean, Phys. Rev. **102**, 1413 (1956).
2. B. Dieny, J. Magn. Magn. Mater. **136**, 335 (1994).
3. J. Nogués and I. K. Schuller, J. Magn. Magn. Mater. **192**, 203 (1999).
4. C. Tsang, T. Lin, S. MacDonald, M. Pinarbasi, N. Robertson, H. Santini and P. Arnett, IEEE Trans/ Magn. **33**, 2866 (1997).
5. R. H. Kovh, G. Grinstein, G. A. Keefe, Y. Lu, P. L. Trouilloud, W. J. Gallagher and S. S. P. Parkin, Phys. Rev. Lett. **84**, 5419 (2000)

6. J. Nogués, J. Sort, V. Langlais, V. Skumryev, S. Surinach, J. S. Munoz, and M. D. Baró, Phys. Rep. **422**, 65 (2005).

7. V. Skumryev, S. Stoyanov, Y. Zhang, G. hadjipanayis, D. Givord and J. Nogues, Nature **423**, 850 (2003).

8. J. A. De Toro, J. P. Andrés, J. A. González, P. Muñiz, T. Muñoz, P. S. Normile, and J. M. Riveiro, Phys. Rev. B **73**, 094449 (2006).

9. S. K. Mishra, F. Radu, H. A. Dürr, and W. Eberhardt, Phys. Rev. Lett. **102**, 177208 (2009).

10. Kentaro Takano, R. H. Kodama, A. E. Berkowitz, W. Cao, and G. Thomas, Phys. Rev. Lett. **79**, 1130(1997).

11. P. Kappenberger, S. Martin, Y. Pellmont, H. J. Hug, J. B. Kortright, O. Hellwig, and Eric E. Fullerton, Phys. Rev. Lett. **91**, 267202 (2003).

12. M. Gibert, P. Zubko, R. Scherwitzl, J. Íñiguez, and J. M. Triscone, Nature Mater. **11**, 195 (2012).

13. Z. M. Tian, S. L. Yuan, S. Y. Yin, L. Liu, J. H. He, H. N. Duan, P. Li, and C. H. Wang, Appl. Phys. Lett. **93**, 222505 (2008).

14. S. Giri, M. Patra and S. J. Majumdar, J. Phys.: Condens. Matter. **23**, 073201 (2011).

15. B. M. Wang, Y. Liu, P. Ren, B. Xia, K. B. Ruan, J. B. Yi, J. Ding, X. G. Li, and L. Wang, Phys. Rev. Lett. 106, 077203 (2011).

16. T. Maity, S. Goswami, D. Bhattacharya, and S. Roy, Phys. Rev. Lett. **110**,107201 (2013).

17. A. K. Nayak, M. Nicklas, S. Chadov, P. Khuntia, C. Shekhar, A. Kalache, M. Baenitz, Y. Skourski, V. K. Guduru, A. Puri, U. Zeitler, J. M. D. Coey & C. Felser, Nature Mater. **14**, 679 (2015).
18. X. D. Tang, W. H. Wang, W. Zhu, E. K. Liu, G. H. Wu, F. B. Meng, H. Y. Liu , and H. Z. Luo, Appl. Phys. Lett. **97**, 242513 (2010).
19. J. K. Murthy and A. Venimadhav, Appl. Phys. Lett. **103**, 252410 (2013).
20. Y. Sun, J. Z. Cong, Y. S. Chai, L. Q. Yan, Y. L. Zhao, S. G. Wang, W. Ning and Y. H. Zhang, Appl. Phys. Lett. **102**, 172406 (2013).
21. M. Sagawa, S. Fujimura, N. Togawa, H. Yamamoto and Y. Matsuura, J. Appl. Phys. **55**, 2083 (1984).
22. V. K. Pecharsky and K. A. Gschneidner Jr., Phys. Rev. Lett. **78** , 4494 (1997).
23. A. E. Clark, in *Ferromagnetic Materials*, edited by E. P. Wohlfahrt (North-Holland, Amsterdam, 1980).
24. J. Deportes, D. Givord, R. Lemaire and H. Nagai, Physica B+C, **86**, 69 (1977).
25. Y. C. Yang, B. Kebe, W. J. James, J. Deportes, and W. Yelon, J. Appl. Phys. **52**, 2077 (1981).
26. M. Morales, M. Bacmann, P. Wolfers, D. Fruchart and B. Ouladdiaf, Phys. Rev. B **64**, 144426 (2001).
27. C. Pique, E. Abad, J. A. Blanco, R. Burriel and M. T. Fernández-Díaz, Phys. Rev. B **71**, 174422 (2005).
28. D. Niebieskikwiat and M. B. Salamon, Phys. Rev. B **72**, 174422 (2005).
29. Y. Tang, Y. Sun, and Z. Cheng, J. Appl. Phys. **100**, 023914 (2006).
30. D. Paccard, C. Schlenker, O. Massenet, R. Montmory and A. Yelon, Phys.

Status Solidi B **16** , 301( 1966).

31. C. Binek, Phys. Rev. B **70**, 014421 (2004).

32. T. Gredig, I. N. Krivorotov, E. D. Dahlberg, Phys. Rev. B 74,094431(2006).

33. W. H. Meikejoh, J. Appl. Phys. 33(3), 1328 (1962).

34. M. D. Stiles and R. D. McMichael, Phys. Rev. B 60, 12950 (1999).

**Figure Captions**

**FIG. 1** The magnetic phases diagram of the $YMn_{12-x}Fe_x$ ($6.0 \leq x \leq 8.8$) under a magnetic field of 1 kOe. $T_C$ (blue circle) represents the Curie temperature, $T_N$ (red triangle) is the Néel temperature, and $T_f$ (black square) is the temperature corresponding the bifurcation point in the zero field cooling and field cooling magnetization curves. PM, AFM, FM and SG are represent paramagnetic, antiferromagnetic, ferromagnetic and spin glass phases, respectively.

**FIG. 2** (a) *M-T* curves of $YMn_{4.4}Fe_{7.6}$ alloy measured under $H$ = 100 Oe after ZFC and FC, (b) *M-H* hysteresis loops at 5 K for $YMn_{4.4}Fe_{7.6}$ under different cooling fields, (c) The dependence of the EB on the cooling magnetic field $H_{\mathrm{cool}}$ at 5 K. (d) The real part $\chi'$ of ac susceptibility curve of $YMn_{4.4}Fe_{7.6}$ at different frequencies with ac magnetic field of 10 Oe after ZFC from 300K. The inset shows the enlarged scale around 130K.

**FIG. 3** The hysteresis loops of $YMn_{4.4}Fe_{7.6}$ after field cooling under 10 kOe at temperatures of (a) 5-28 K (the inset shows the training effect $H_E$ at 5K) and (b) 28-100K.

**FIG. 4** Temperature dependence of (a) $H_L$, $H_R$ and (b) $H_C$, $H_E$ of $YMn_{4.4}Fe_{7.6}$ alloy at 5 K after FC with 10 kOe.

**FIG. 5** (a) The schematic magnetic structure of the $YMn_{4.4}Fe_{7.6}$ compound (top view along c axis). (b) Schematic diagram of angles involved in the theoretical model. (c) The calculated magnetic hysteresis loops as a function of $K_{AFM}$ . (c) The obtained $H_L$, $H_R$, $H_C$ and $H_E$ as functions of $K_{AFM}^{-1}$ .

Table I: The magnetic properties of the $YFe_xMn_{12-x}$ alloys (6.0 < x < 8.8).

| | $H_C$ (Oe) | $H_{EB}$ (Oe) | $H_C$ (Oe) | $H_{EB}$ (Oe) |
|---|---|---|---|---|
| Cooling field | 0 Oe | 0 Oe | 1 kOe | 1 kOe |
| $YFe_{8.8}Mn_{3.2}$ | -- | -- | 164 | 58 |
| $YFe_{8.4}Mn_{3.6}$ | 132 | 0 | 114 | 663 |
| $YFe_{8.2}Mn_{3.8}$ | 222 | 0 | 60 | 1483 |
| $YFe_{8.0}Mn_{4.0}$ | 979 | 0 | 409 | 1934 |
| $YFe_{7.8}Mn_{4.2}$ | 823 | 0 | 284 | 2489 |
| $YFe_{7.6}Mn_{4.4}$ | 85 | 0 | 81 | 5956 |
| $YFe_{7.4}Mn_{4.6}$ | 27 | 0 | 23 | 5180 |
| $YFe_{7.0}Mn_{5.0}$ | 9889 | 2802 | 9821 | 2876 |
| $YFe_{6.6}Mn_{5.4}$ | 7429 | 641 | 7496 | 660 |
| $YFe_{6.0}Mn_{6.0}$ | -- | -- | 472.5 | 76.5 |

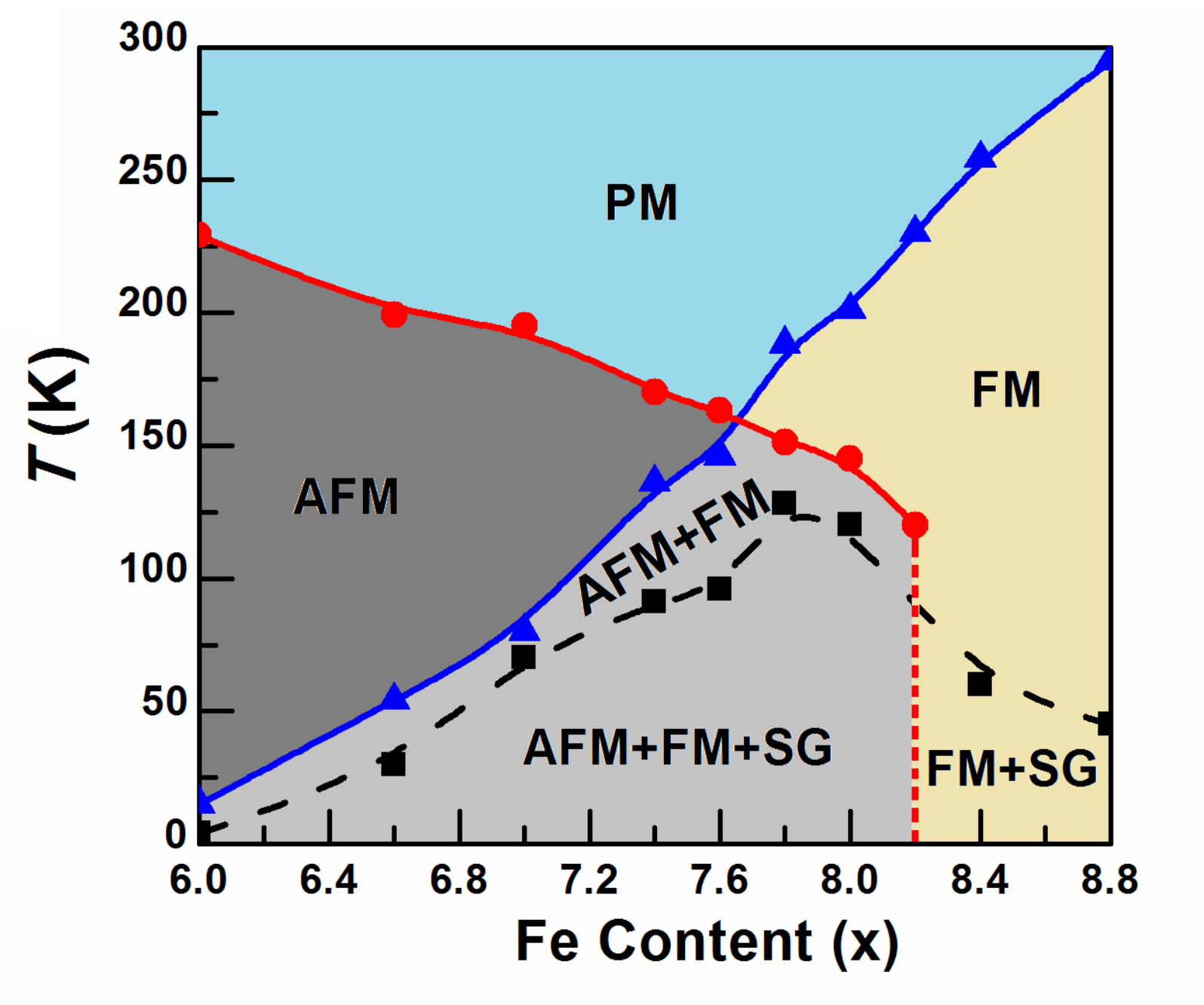

Figure 1. Xia et al.

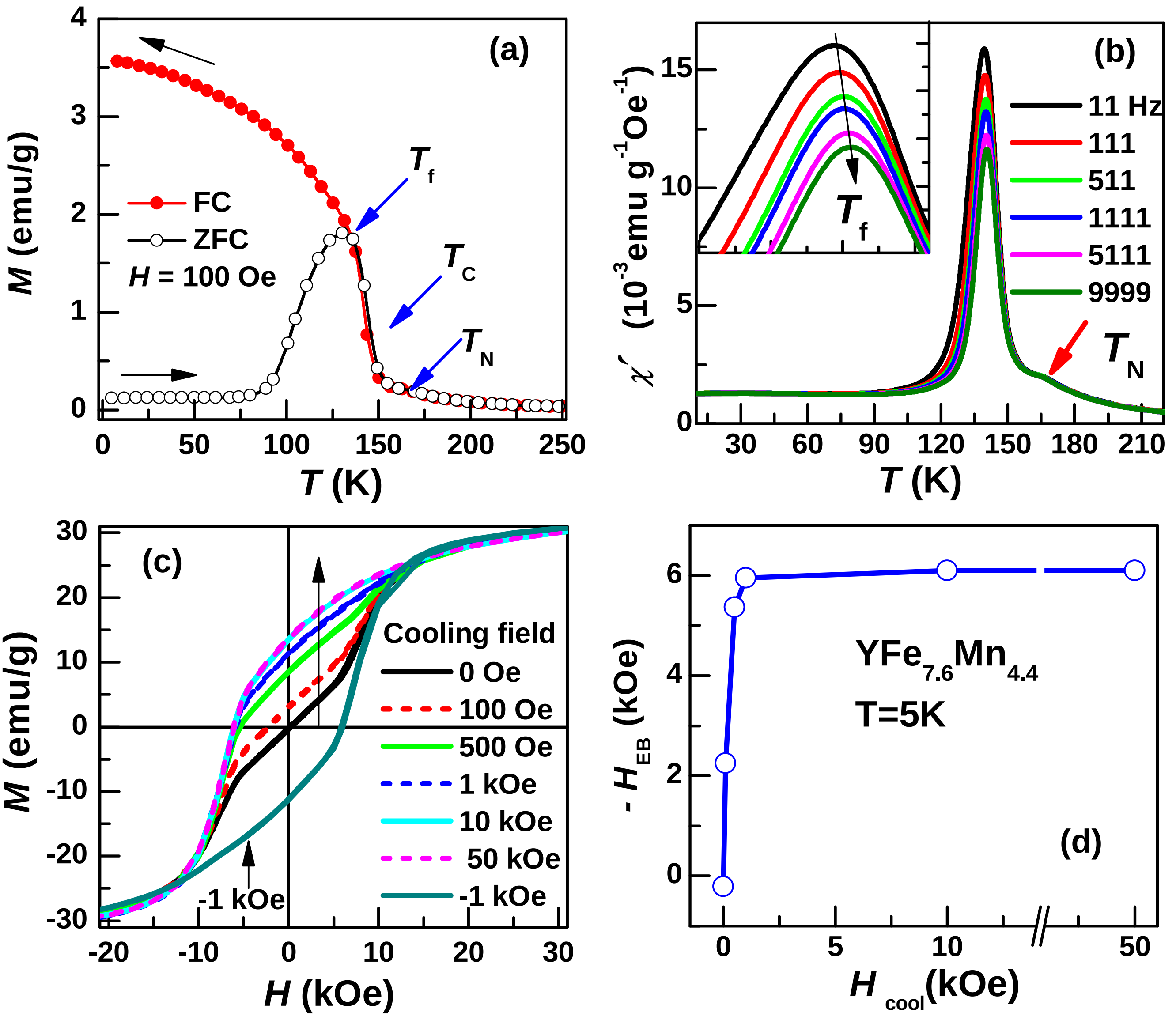

Figure 2. Xia et al.

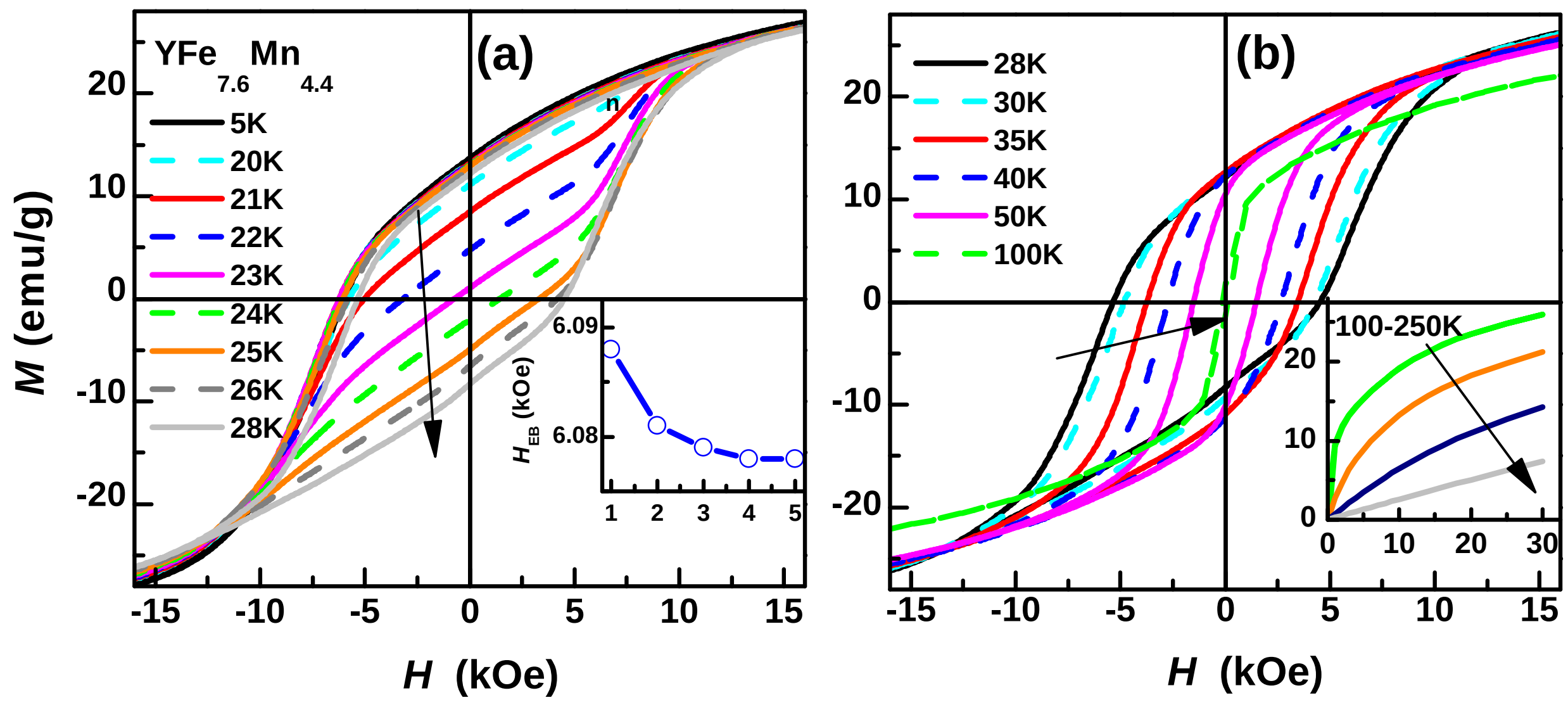

Figure 3. Xia et al.

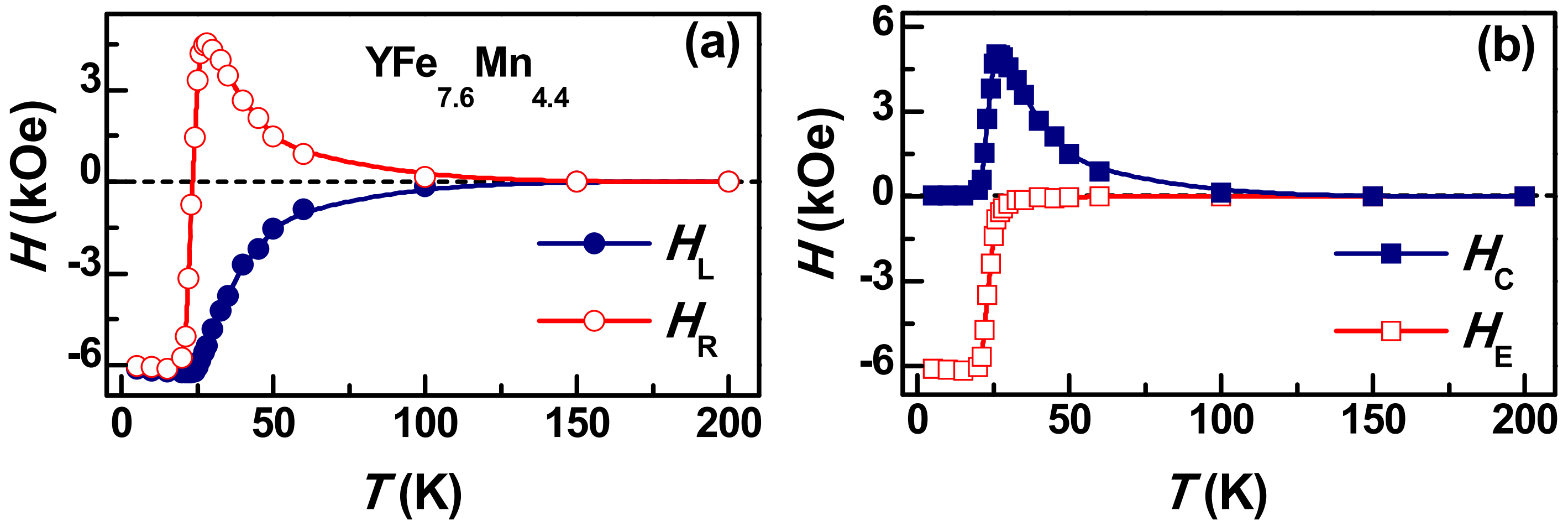

Figure 4. Xia et al.

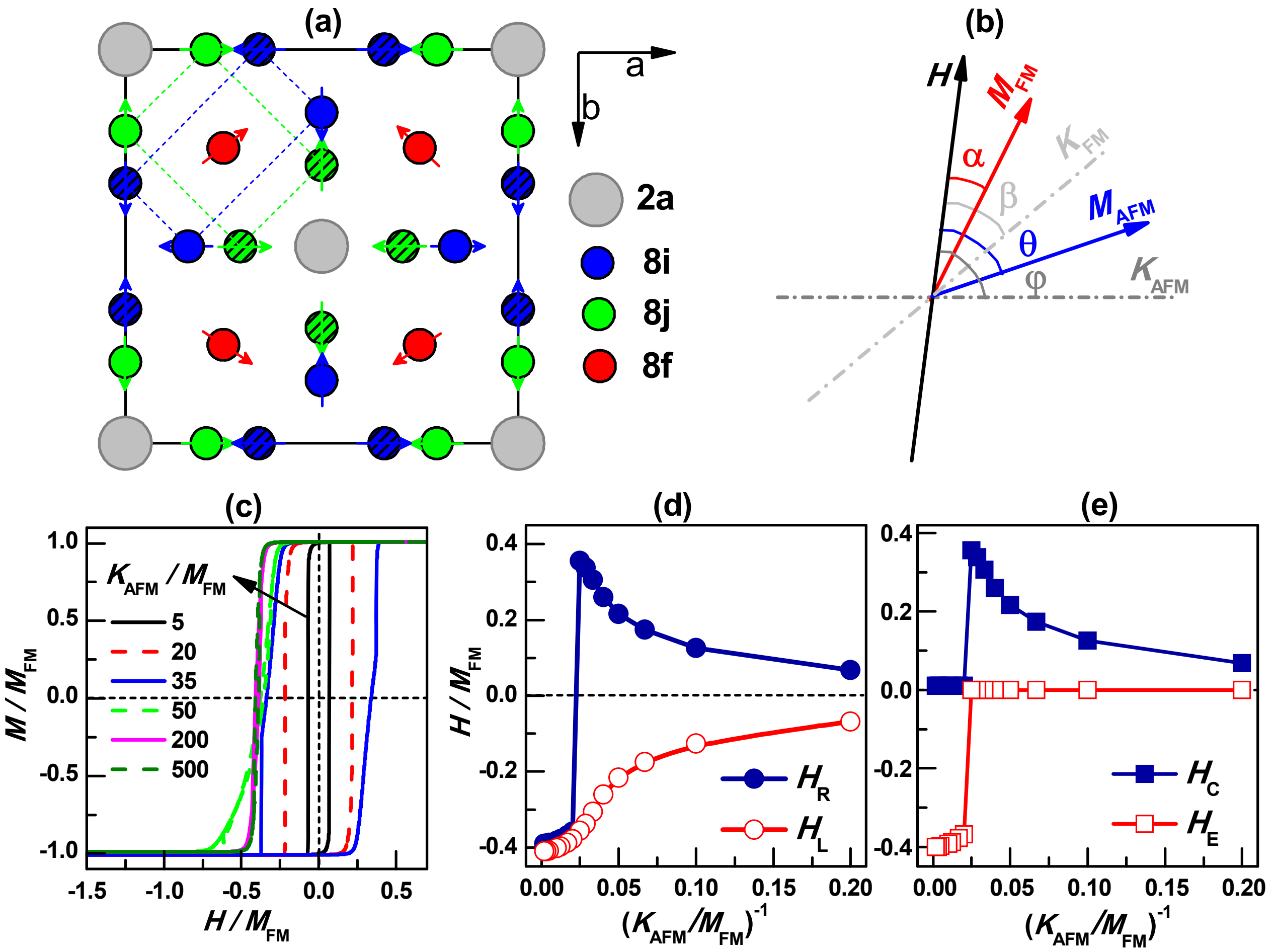

Figure 5. Xia et al.